\documentclass[twocolumn]{jjap2}
\usepackage{txfonts}

\title{Acceleration of amyloid fibril formation by multichannel sonochemical reactor}

\author{Kentaro Noi$^{1}$, Kichitaro Nakajima$^{2}$, Keiichi Yamaguchi$^{2}$, Masatomo So$^{3}$, Kensuke Ikenaka$^{4}$, Hideki Mochizuki$^{4}$, Yuji Goto$^{2}$, and Hirotsugu Ogi$^{5}$\thanks{E-mail: ogi@prec.eng.osaka-u.ac.jp}}

\inst{
$^{1}$Institute for NanoScience. Design, Osaka University, Toyonaka, Osaka 560-8531, Japan\\
$^{2}$Global Center for Medical Engineering and Informatics, Osaka University, Suita, Osaka 565-0871, Japan\\
$^{3}$Astbury Centre for Structural Molecular Biology, University of Leeds, Leeds LS2 9JT, U.K.\\
$^{4}$Department of Neurology, Graduate School of Medicine, Osaka University, Suita, Osaka 565-0871, Japan\\
$^{5}$Graduate School of Engineering, Osaka University, Suita, Osaka 565-0871, Japan\\
}

\abst{Formation of amyloid fibrils of various amyloidogenic proteins is dramatically enhanced by ultrasound irradiation. For applying this phenomenon to the study of protein aggregation science and diagnosis of neurodegenerative diseases, a multichannel ultrasound irradiation system with individually adjustable ultrasound-irradiation conditions is necessary.  Here, we develop a sonochemical reaction system, where an ultrasonic transducer is placed in each well of a 96-well microplate to perform ultrasonic irradiation of sample solutions under various conditions with high reproducibility, and applied it for studying amyloid-fibril formation of amyloid $\beta$, $\alpha$-synuclein, $\beta$2-microglobulin, and lysozyme. The results clearly show that our instrument is superior to conventional shaking method in terms of degree of acceleration and reproducibility of fibril formation reaction. The acceleration degree is controllable by controlling the driving voltage applied to each transducer.  We have thus succeeded in developing a useful tool for the study of amyloid fibril formation in various proteins.  
}

\begin{document}
\maketitle

\section{Introduction}
In recent years, ultrasound technology has made remarkable progress in medical applications, and is often used for imaging\cite{Yabata, Prastika}, biosensors\cite{Zhou1, Kato, Noi2021}, and gas sensors\cite{Zhou2,Zhou3}, and the ultrasound frequencies used in these applications are generally in the megahertz band. However, it has become clear that even low-frequency ultrasound waves with a frequency of several tens of kilohertz can interact with small proteins and dramatically affect their reactions as described below.

Amyloid fibril is misfolded aggregate of specific protein, which shows a highly ordered, $\beta$-sheet-rich, needle-like structure\cite{Chiti,Eisenberg}. When these amyloid fibrils are deposited inside and outside of nerve cells, they are involved in the development of neurodegenerative diseases \cite{Hardy, Berthelot, Hoshi}. The proteins that make up amyloid fibrils are different in each neurodegenerative disease, such as amyloid $\beta$ (A$\beta$) peptide and Tau protein in Alzheimer's disease and $\alpha$-synuclein in Parkinson's disease.\cite{Stephan, Wischik, Kramer, Masters, Spillantini, Spillantini2}. Interestingly, a common reaction process has been observed in the formation of amyloid fibrils regardless of the type of constituent protein: The misfolded proteins form the nucleus, followed by an elongation reaction in which monomeric proteins are adsorbed to the ends of the nuclei and fibrils to elongate the fibrils.\cite{Arosio,Ferrone,Lee} Another common feature is that the energy barrier for the nucleation reaction is very high\cite{Jerrett, Nakajima1}, and the fibril formation reaction takes a long time. In order to promote amyloid fibril formation, considerably higher concentrations of protein compared to biological concentrations and agitation such as stirring and rotation have been used\cite{Giehm}.  The fibril formation behavior can be monitored by a fluorometric analysis using thioflavin T (ThT), which specifically binds to the $\beta$-sheet structure in amyloid fibrils and emits high-intensity fluorescence\cite{Naiki}. An increase in the ThT fluorescence intensity therefore indicates the increase in formed fibrils.

Ultrasound irradiation is a beneficial method to significantly promote the nucleation of amyloid fibrils. It repeatedly creates positive and negative pressures in the solution.The mechanism of accelerated nucleation reaction is explained by the interaction between cavitation bubbles generated by ultrasound wave and dissolved proteins. The cavitation bubbles grow under negative pressure, and a large amount of protein molecules are trapped on the surfaces of the bubbles by their hydrophobic sites. Then, when the acoustic pressure switches to positive pressure, the bubbles contract and collapse in a short time, so that the adsorbed protein molecules are collected at contraction points of the bubbles and condensed locally, and at the same time, a transient temperature increase due to adiabatic compression of the gas in the bubble occurs, which accelerates the nucleation reaction.\cite{Uesugi, Nakajima1,Nakajima2}. 

We have developed a high-throughput fluorescence spectroscopy system that can monitor the formation process of amyloid fibrils while performing ultrasound stimulation and named it "HANABI" (Handai amyloid burst inducer).\cite{So, Umemoto, Yoshimura1}. In this system, a 96-well microplate containing protein solutions is placed near the water surface in a water bath, and three ultrasound generators simultaneously irradiate the microplate from below in the water bath. We applied the HANABI system to various proteins and found that it was an effective system for studying their fibril formation behavior\cite{So,Chatani, Yagi,Muta,Yagi2,Kakuda}. However, there are several problems with the HANABI system: The powerful ultrasonic waves emitted from the three ultrasonic generators fixed at the bottom of the water bath generate a large number of cavitation bubbles in the bath, which scatter the ultrasonic waves, resulting in uneven ultrasonic energy input to each well of the microplate. In addition, the temperature in the bath fluctuates due to ultrasound, resulting in poor reproducibility. 

To solve these problems, we developed a multichannel ultrasonic sonochemical reactor with fluorescence measurement system, in which the water bath was removed and one transducer was placed in each well of a 96-well microplate (Fig. 1\cite{Noi}). As a result, we successfully detected the amyloid fibril formation of $\beta$2-microglobulin  ($\beta$2m)  and the low concentration seed of $\beta$2m\cite{NakajimaACS,NakajimaUS}. We name this improved HANABI system "HANABI-2000". In the previous study\cite{Noi}, we have shown that the developed multichannel ultrasound reactor is capable of inducing fibrillation reactions of A$\beta_{1-40}$ and $\alpha$-synuclein solutions.  Here, the applicability of the developed system to amyloid assay measurements was investigated in more detail, not only for A$\beta_{1-40}$ and $\alpha$-synuclein, but also for $\beta$2m and lysozyme from egg white.  We confirmed that the multichannel sonochemical reactor can significantly promote the formation of amyloid fibrils of these proteins, and that the reaction rate of each well can be controlled by adjusting the ultrasound irradiation conditions individually.

\section{Experimental methods and instruments}
\subsection{ Materials and protein preparation}
Lyophilized-powder A$\beta_{1-40}$ peptides were purchased from Peptide Institute (No. 4307-v). Recombinant human $\beta$2m protein with an additional methionine residue at the N terminus and recombinant human $\alpha$-synuclein were expressed in E. coli and purified as described elsewhere \cite{Chiti,Chiba,Yagi3}. Lysozyme, dimethyl sulfoxide (DMSO), phosphate-buffer saline (PBS), NaCl, and Thioflavin-T (ThT) were purchased from Wako Pure Chemical Industries Ltd.

The powder of A$\beta_{1-40}$ was dissolved in DMSO, and the solution was diluted with 100 mM PBS mixed with NaCl and ThT to obtain a final concentration of 5 $\mu$M of A$\beta_{1-40}$, 100 mM of NaCl, and 5 $\mu$M of ThT at pH 7.4. (The volume ratio of DMSO to PBS is 1:4.) 

Lyophilized powdered $\alpha$-synuclein was dissolved in 0.8 M phosphate buffer (pH 6.9) containing ThT to obtain a solution with a final concentration of 7.0 $\mu$M of $\alpha$-synuclein with 5 $\mu$M ThT.  

Lyophilized powdered $\beta$2m was dissolved in 10 mM HCl (pH 2.0) containing NaCl and ThT to obtain a final concentration of 8.5 $\mu$M of $\beta$2m with 100 mM NaCl and 5 $\mu$M ThT. 

Powder lysozyme from egg white (No. 20841-54, Nacalai Tesque Inc.) was dissolved by 3 M guanidinium chloride containing HCl and ThT to obtain a final concentration of 35 $\mu$M of lysozyme with 10 mM NaCl and 5 $\mu$M ThT.	

\begin{figure}
\begin{center}
\includegraphics[width=85mm]{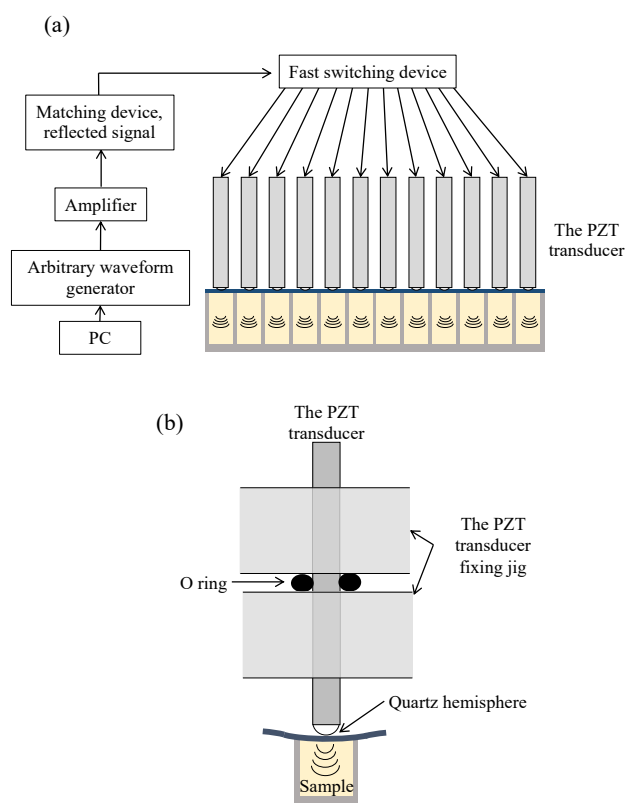}
\end{center}
\caption{(a) Schematic of a multichannel sonochemical reactor (HANABI-2000). Each piezoelectric lead zirconate titanate (PZT) transducer with dimensions of 4.24$\times$4.24$\times$54 mm$^3$ is sequentially fed a 0.3-ms-long burst signal by a fast switching device. (b) Schematic of a single PZT transducer and one well of a 96-well plate.}
\label{Fig1.eps}
\end{figure}

\subsection{Multiplechannel sonochemical reactor}
We have developed a spectroscopic multichannel sonochemical reactor, as shown in Fig. 1\cite{Noi}.  A single piezoelectric lead zirconate titanate (PZT) transducer is placed in each well of a 96-well microplate and it irradiates the sample solution in the well through a thin ($<\sim$0.1 mm) plastic sheet with ultrasound, and the fluorescence measurement is performed from the back surface.   The PZT transducers are driven in turn by a 0.1 ms burst signal via a fast switching device. The amplitude and frequency of the burst signal can be controlled individually for each well. Typical amplitude and frequency of the burst signal are 100 V and 29 kHz, respectively.

The sample solutions were pipetted into a 96-well microplate. Each well was filled with the sample solution and sealed with the 0.1 mm thick plastic film so as to prevent air from entering.  The 96-well microplate was then set in the multichannel sonochemical reactor. A quartz hemisphere was bonded to the end face of each PZT transducer, which makes point contact with the plastic film, introducing ultrasonic waves. The PZT transducer was supported by an O-ring at the node of the resonant vibration. In each well, the sample solution was sonicated for 0.1 ms followed by 8 s of quiescence.  This set of 0.1-ms irradiation and 8-s quiescence was repeated for several tens of hours in all wells, acquiring the fluorescence intensity every 10 min.  The amount of the formed amyloid fibrils was evaluated by the fluorescence intensity of ThT, where the ThT molecule was excited at 450 nm, and its fluorescence at 485 nm was detected.

\subsection{AFM measurement}
The morphology of the aggregates formed in the solution was observed by atomic force microscopy (AFM) using an instrument produced by Shimadzu Corporation (SPM A9600).  The tapping mode was used with a silicon cantilever (stiffness: 40 N/m, resonant frequency: 300 kHz). The ultrasonically irradiated solution was dropped onto the mica with the new surface and incubated for 15 min. Then, it was washed with 100 $\mu$l of ultrapure water and dried before AFM observation.

\begin{figure}
\begin{center}
\includegraphics[width=85mm]{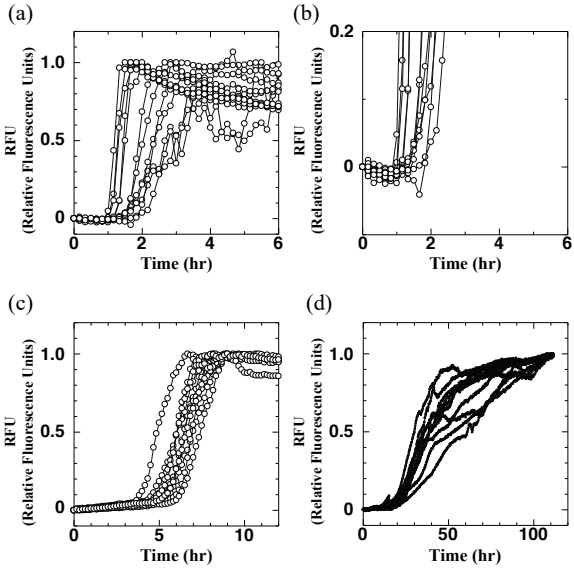}
\end{center}
\caption{Changes in ThT fluorescence intensity for A$\beta_{1-40}$ peptide (a),(b) measured by the multichannel sohochemical reactor, (c) under shaking, and (d) under quiescent. In (b), data extracted from the initial stage of the reaction in (a) is shown. }
\label{Fig2.eps}
\end{figure}

\section{Results and Discussion} 
Figure 2(a) shows the time course of ThT fluorescence intensity of A$\beta_{1-40}$ solution measured by the HANABI-2000 system. These are the results of simultaneous measurements of 12 wells in one microplate.  Although there seems to be some variation in the change in fluorescence intensity, the variation is small enough when we focus on the initial rise time of the fluorescence intensity (Fig. 2(b)). The rapid increase in the fluorescence intensity corresponds to the onset of amyloid fibril formation.  For comparison,  changes in the ThT fluorescence intensity were measured under shaking and quiescent conditions (Figs. 2(c) and (d), respectively).  When the threshold of the lag time to amyloid formation is set to 0.1 of the maximum ThT intensity, the average values of the lag time obtained in the multichannel sonochemical reactor, shaking, and quiescent are 1.7, 5.1 and 23 h, respectively.  Therefore, our developed multichannel sonochemical reactor can significantly accelerate the amyloid fibril formation of A$\beta_{1-40}$ peptide. 

\begin{figure}
\begin{center}
\includegraphics[width=80mm]{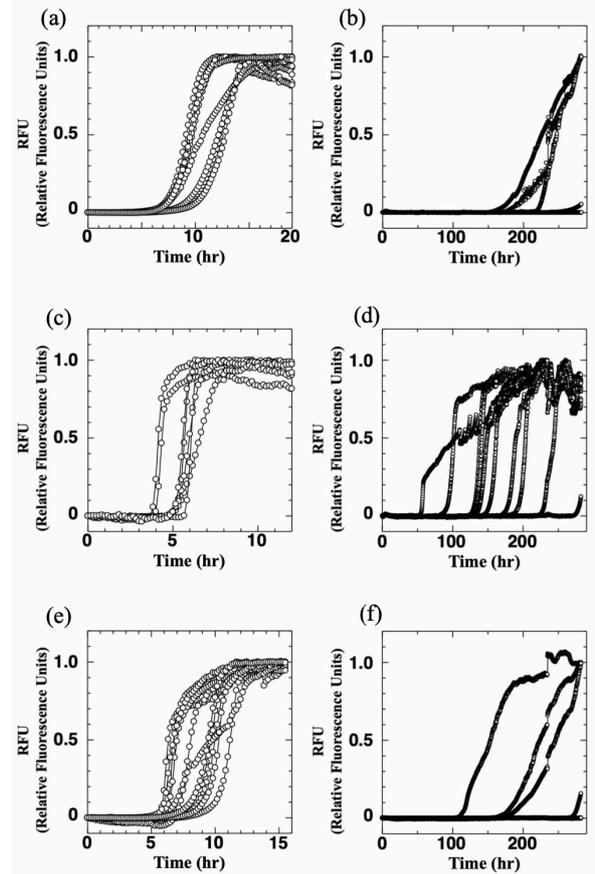}
\end{center}
\caption{Time-course curves of ThT fluorescence intensity for $\alpha$ synuclein ((a) and (b)), $\beta$2m ((c) and (d)), and lysozyme ((e) and (f)).  (a), (c), and (e) are measurements by the multichannel sonochemical reactor; and (b), (d), and (f) are obtained under shaking condition.}
\label{Fig3.eps}
\end{figure}

Figures 3 (a), (c), and (e) show the results of the measurements for $\alpha$-synuclein,  $\beta$2m, and lysozyme, respectively. All protein solutions showed a rapid increase in the ThT fluorescence intensity within 15 h when using our multichannel sonochemical reactor.  We used the coefficient of variation (CV) value as a measure of reproducibility, which is the standard deviation divided by the average value.  Table I shows the average lag time, standard deviation, and the CV value.  It can be seen that the CV values are very low for the measurements for all samples, indicating high reproducibility among different wells. For example, in our previous study with the high-throughput fluorescence spectroscopy system (previous HANABI system), the CV value for $\beta$2m solution was about 20\% at the optimal conditions under plate movement\cite{Umemoto}, which was reduced to 14\% in the HANABI-2000 instrument. It is also important to note that AFM observation confirmed the formation of amyloid fibrils in all sample solutions as shown in Fig. 4. Although it is reported that ultrasound irradiation causes fragmentation of amyloid fibrils\cite{Chatani2}, sufficiently long amyloid fibrils up to 5  $\mu$m were observed in the present study. Therefore, the ultrasound irradiation by the multichannel sonochemical reactor developed in this study contributes to the elongation of amyloid fibrils rather than their fragmentation.

On the other hand, under shaking condition, the ThT fluorescence intensity did not increase in some wells even after 280 h (Figs. 3(b), (d), and (f)), and the lag time variation was significantly large (Table I). These results indicate that the HANABI-2000 system developed in this study not only promotes the formation of amyloid fibrils of various proteins, but also achieves high reproducibility.

\begin{table}[tb]
\begin{center}
\caption{Average (Ave), standard deviation (SD), and coefficient of variation (CV) of lag time for amyloid fibril formation A$\beta_{1-40}$, $\alpha$ synuclein, $\beta$2m, and lysozyme observed by the multichannel sonochemical reactor (HANABI-2000), shaking, and quiescent. There were wells in which ThT intensity failed to increase under shaking and quiescent conditions, and the lag time was considered as the time at the end of the measurement (280 h) for them.
}
\label{parameter}
\begin{tabular}{ccccc} \hline\hline
 &  & Average (h)  & SD (h)  & CV (\%)\\\hline
A$\beta_{1-40}$ & HANABI-2000 & 1.65  & 0.42  & 25.2\\
 & shaking & 5.09  & 0.59  & 11.5\\
 & quiescent  & 23.3  & 3.73  & 16.0\\
 \hline
$\alpha$ synuclein & HANABI-2000 & 8.71  & 1.20  & 13.8\\
 & shaking & 238 & 66.3 & 27.9\\
\hline
$\beta$2m & HANABI-2000 & 5.12  & 0.73  & 14.3\\
 & shaking & 160 & 65.2 & 40.6\\
\hline
Lysozyme & HANABI-2000 & 7.64  & 1.36  & 18.0\\
 & shaking & 248 & 57.8 & 23.3\\
 \hline\hline
 \end{tabular}
\end{center}
\end{table}

\begin{figure}
\begin{center}
\includegraphics[width=90mm]{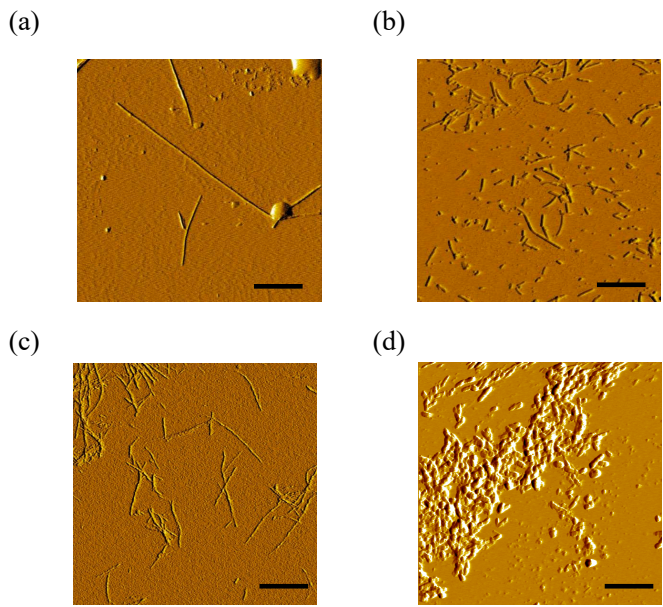}
\end{center}
\caption{AFM images of (a) A$\beta_{1-40}$, (b) $\alpha$ synuclein, (c) $\beta$2m, and (d) lysozyme, respectively, after ultrasonic irradiation by the HANABI-2000 system. The scale bars indicate 1 $\mu$m.}
\label{Fig4.eps}
\end{figure}

\begin{figure}
\begin{center}
\includegraphics[width=90mm]{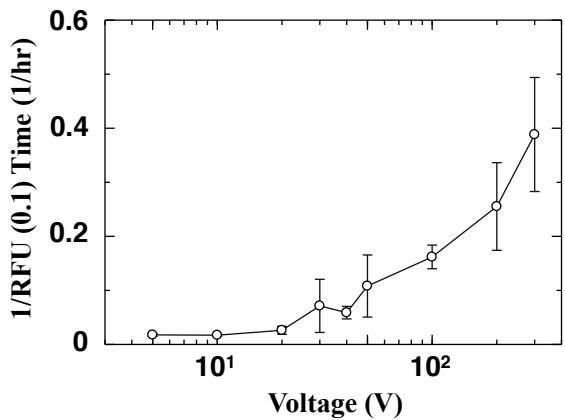}
\end{center}
\caption{Relationship between voltage applied to piezoelectric transducer and inverse of lag time in fibril formation reaction of A$\beta_{1-40}$ peptide.}
\label{Fig5.eps}
\end{figure}

Finally, we investigated the effect of the voltage applied to the PZT transducer on the fibril formation reaction. Figure 5 shows the relationship between the inverse of the lag time and the applied voltage for amyloid fibril formation using A$\beta_{1-40}$ peptide. The result indicates that as the applied voltage increases, the lag time becomes shorter, and the fibril formation reaction can be controlled by adjusting the applied voltage.  This characteristic is very important in the diagnosis of neurodegenerative diseases. Recently, a method has been proposed to amplify amyloid fibrils by adding body fluids, such as cerebrospinal fluid, to a protein solution that causes neurodegenerative diseases, using the seeding effect to diagnose the disease. This is because the presence of seeds indicates the onset of the disease and promotes the formation of amyloid fibrils. However, this method has the problem that it takes a long time, about 300 h, before amyloid fibrils are detected\cite{Soto}. Since ultrasound irradiation is expected to promote the fibril growth reaction, it is expected to speed up this type of diagnosis. However, at the same time, it significantly accelerates the primary nucleation reaction that fibrillates monomers, and if the ultrasound sound pressure is too high, it will be difficult to distinguish between the presence or absence of the seed. If the sound pressure is lowered, fibril formation will take longer. Therefore, it is considered that there is an optimal ultrasound pressure for the diagnosis of each disease, and the multichannel sonochemical system developed here, which can irradiate the solution in each well at different pressure, will highly contribute to the early stage diagnosis of neurodegenerative diseases.

\section{Conclusion} 
We throughout examined the usefulness of our developed multichannel sonochemical reactor, HANABI-2000. It is capable of highly accelerating amyloid fibril formation of various proteins with high reproducibility among many wells of a microplate. The amyloid fibril formation sometimes takes more than 300 h under the traditional shaking condition, but our multichannel sonochemical reactor has succeeded in shorting the assay time to less than 20 h. In this system, ultrasonic-irradiation conditions (amplitude, frequency, duration, and so on) on each PZT transducer can be easily adjusted, allowing simultaneous measurement of different conditions for various proteins. Therefore, the multichannel sonochemical reactor will be a powerful tool for studying amyloid fibril formation and diagnosis of neurodegenerative diseases.


\begin{thebibliography}{9}
\bibitem{Yabata}H. Yabata, S. Umemura, and S. Yoshizawa: Jpn. J. Appl. Phys. 60 (2021) SDDE23. 
\bibitem{Prastika}E. B. Prastika, A. Imori, T. Kawashima, Y. Murakami, N. Hozumi, S. Yoshida, R. Nagaoka, and K. Kobayashi: Jpn. J. Appl. Phys. 60 (2021) SDDE22. 
\bibitem{Zhou1}L. Zhou, F. Kato, and H. Ogi: Jpn. J. Appl. Phys. 60 (2021) SDDB03. 
\bibitem{Kato}F. Kato, Y. Sato, H. Ato, H. Kuwabara, Y. Kobayashi, K. Nakamura, N. Masumoto, H. Noguchi, and H. Ogi: Jpn. J. Appl. Phys. 60 (2021) SDDC01 (2021).
\bibitem{Noi2021}K. Noi, K. Ikenaka, H. Mochizuki, Y. Goto, and H. Ogi: Anal. Chem. 93 (2021) 11176 (2021).
\bibitem{Zhou2}L. Zhou, N. Nakamura, A. Nagakubo, and H. Ogi: Jpn. J. Appl. Phys. 59 (2020) SKKB02. 
\bibitem{Zhou3}L. Zhou, F. Kato, N. Nakamura, Y. Oshikane, A. Nagakubo, and H. Ogi: Sens. Actuat. B. 334 (2021) 129651.

\bibitem{Chiti}F. Chiti and C.M. Dobson: Annu Rev Biochem. 86 (2017)27. 
\bibitem{Eisenberg}D. Eisenberg and M. Jucker: Cell. 148 6 (2012)1188. 
\bibitem{Hardy}J.A. Hardy and G.A. Higgins: Science. 256 [5054](1992)184. 
\bibitem{Berthelot}K. Berthelot, C. Cullin and S. Lecomte: Biochimie. 95 [1](2013)12. 
\bibitem{Hoshi}M. Hoshi, M. Sato, S. Matsumoto, A. Noguchi, K. Yasutake, N. Yoshida and K. Sato: Proc Natl Acad Sci U S A. 100 [11](2003)6370. 
\bibitem{Stephan}A. St\'{e}phan, S. Laroche and S. Davis: J Neurosci. 21 [15](2001)5703. 
\bibitem{Wischik}C.M. Wischik, C.R. Harrington and J.M. Storey: Biochem Pharmacol. 88 [4](2014)529. 
\bibitem{Kramer}M.L. Kramer and W.J. Schulz-Schaeffer: J Neurosci. 27 [6](2007)1405. 
\bibitem{Masters}C.L. Masters, G. Simms, N.A. Weinman, G. Multhaup, B.L. McDonald and K. Beyreuther: Proc Natl Acad Sci U S A. 82 [12](1985)4245. 
\bibitem{Spillantini}M.G. Spillantini, M.L. Schmidt, V.M. Lee, J.Q. Trojanowski, R. Jakes and M. Goedert: Nature. 388 [6645](1997)839. 
\bibitem{Spillantini2}M.G. Spillantini, R.A. Crowther, R. Jakes, M. Hasegawa and M. Goedert: Proc Natl Acad Sci U S A. 95 [11](1998)6469. 
\bibitem{Arosio}P. Arosio, T.P. Knowles and S. Linse: Phys Chem Chem Phys. 17 [12](2015)7606. 
\bibitem{Ferrone}F. Ferrone: Methods Enzymol. 309 (1999)256..
\bibitem{Lee}C.C. Lee, A. Nayak, A. Sethuraman, G. Belfort and G.J. McRae: Biophys J. 92 [10](2007)3448. 
\bibitem{Jerrett}J. T. Jerrett and P. T. Lansbury Jr.: Cell, 73 (1993) 1055. 
\bibitem{Nakajima1}K. Nakajima, H. Ogi, K. Adachi, K. Noi, M. Hirao, H. Yagi and Y. Goto: Sci Rep. 6 (2016)22015. 


\bibitem{Giehm}L. Giehm and D.E. Otzen: Anal Biochem. 400 [2](2010)270. 
\bibitem{Naiki}H. Naiki, K. Higuchi, M. Hosokawa, and T. Takeda: Anal. Biochem. 177 (1989) 244.

\bibitem{Uesugi}K. Uesugi, H. Ogi, M. Fukushima, M. So, H. Yagi, Y. Goto, and M. Hirao: Jpn. J. Appl. Phys. 52 (2013) 07HE10.

\bibitem{Nakajima2}K. Nakajima, D. Nishioka, M. Hirao, M. So, Y. Goto and H. Ogi: Ultrason Sonochem. 36 (2017)206. 
\bibitem{So}M. So, H. Yagi, K. Sakurai, H. Ogi, H. Naiki and Y. Goto: J Mol Biol. 412 [4](2011)568. 
\bibitem{Umemoto}A. Umemoto, H. Yagi, M. So and Y. Goto: J Biol Chem. 289 [39](2014)27290. 
\bibitem{Yoshimura1}Y. Yoshimura, M. So, H. Yagi, and Y. Goto:Jap. J. Appl. Phys. 52 (2013) 07HA01.

\bibitem{Chatani}E. Chatani, H. Yagi, H. Naiki and Y. Goto: J Biol Chem. 287 [27](2012)22827. 
\bibitem{Yagi}H. Yagi, K. Hasegawa, Y. Yoshimura and Y. Goto: Biochim Biophys Acta. 1834 [12](2013)2480. 
\bibitem{Muta}H. Muta, Y.H. Lee, J. Kardos, Y. Lin, H. Yagi and Y. Goto: J Biol Chem. 289 [26](2014)18228. 
\bibitem{Yagi2}H. Yagi, A. Mizuno, M. So, M. Hirano, M. Adachi, Y. Akazawa-Ogawa, Y. Hagihara, T. Ikenoue, Y.H. Lee, Y. Kawata and Y. Goto: Biochim Biophys Acta. 1854 [3](2015)209. 
\bibitem{Kakuda}K. Kakuda, K. Ikenaka, K. Araki, M. So, C. Aguirre, Y. Kajiyama, K. Konaka, K. Noi, K. Baba, H. Tsuda, S. Nagano, T. Ohmichi, Y. Nagai, T. Tokuda, O.M.A. El-Agnaf, H. Ogi, Y. Goto and H. Mochizuki: Sci Rep. 9 [1](2019)6001. 
\bibitem{Noi}K. Noi, K. Nakajima, K. Yamaguchi, M. So, K. Ikenaka, H. Mochizuki, Y. Goto, H. Ogi, Proc. 42th Symp. Ultrasonic Electronics 42 (2021) 1Pa4-5
\bibitem{NakajimaACS}K. Nakajima, H. Toda, K. Yamaguchi, M. So, K. Ikenaka, H. Mochizuki, Y. Goto and H. Ogi: ACS Chem Neurosci. 12 [18](2021)3456. 
\bibitem{NakajimaUS}K. Nakajima, K. Noi, K. Yamaguchi, M. So, K. Ikenaka, H. Mochizuki, H. Ogi and Y. Goto: Ultrason Sonochem. 73 (2021)105508. 
\bibitem{Chiba}T. Chiba, Y. Hagihara, T. Higurashi, K. Hasegawa, H. Naiki and Y. Goto: J Biol Chem. 278 [47](2003)47016. 
\bibitem{Yagi3}H. Yagi, E. Kusaka, K. Hongo, T. Mizobata and Y. Kawata: J Biol Chem. 280 [46](2005)38609. 
\bibitem{Chatani2}E. Chatani, Y. H. Lee, H. Yagi, Y. Yoshimura, H. Naiki, and Y. Goto: Proc. Natl. Acad. Sci. U.S.A. 106, (2009) 11119. 
\bibitem{Soto}M. Shahnawaz, A. Mukherjee, S. Pritzkow, N. Mendez, P. Rabadia, X. Liu, B. Hu, A. Schmeichel, W. Singer, G. Wu, A.-L. Tsai, H. Shirani, K. P. R. Nilsson, P. A. Low, and C. Soto, Nature, 578 (2020) 273.

\end{thebibliography}
\end{document}